\documentclass[final, conference]{IEEEtran}
\usepackage[utf8]{inputenc}
\usepackage{graphicx}
\usepackage[noadjust]{cite}
\usepackage{url}
\usepackage{multirow}
\usepackage{footnote}
\usepackage[ dvips, 
  bookmarks = true,
  bookmarksopen = true,
  bookmarksnumbered = true,
  breaklinks = true,
  linktocpage,
  pagebackref = true,
  colorlinks = true,
  linkcolor = blue,
  urlcolor  = blue,
  citecolor = red,
  anchorcolor = green,
  hyperindex = true,
  hyperfigures
  ]{hyperref}
\usepackage[hyphenbreaks]{breakurl}

\title{Duplicate detection methodology\\ for IP network traffic analysis}

\author{
  \authorblockN{Iñaki Ucar, Daniel Morato, Eduardo Magaña and Mikel Izal}

  \authorblockA{
    Dept. of Automatics and Computer Science\\
    Public University of Navarre\\
    Campus Arrosadia, 31006 Pamplona, Spain\\
    \{i.ucar,daniel.morato,eduardo.magana,mikel.izal\}@unavarra.es
  }
}

\begin{document}

\maketitle

\begin{abstract}
Network traffic monitoring systems have to deal with a challenging problem: the traffic capturing process almost invariably produces duplicate packets. In spite of this, and in contrast with other fields, there is no scientific literature addressing it. This paper establishes the theoretical background concerning data duplication in network traffic analysis: generating mechanisms, types of duplicates and their characteristics are described. On this basis, a duplicate detection and removal methodology is proposed. Moreover, an analytical and experimental study is presented, whose results provide a dimensioning rule for this methodology.
\end{abstract}\vspace*{3mm}


\let\thefootnote\relax\footnote{
 \\\textcopyright 2013 IEEE. Personal use of this material is permitted. Permission from IEEE must be obtained for all other uses, in any current or future media, including reprinting/republishing this material for advertising or promotional purposes, creating new collective works, for resale or redistribution to servers or lists, or reuse of any copyrighted component of this work in other works.\\

 \noindent DOI: \hyperref{http://dx.doi.org/10.1109/IWMN.2013.6663796}{}{}{10.1109/IWMN.2013.6663796}
}

\section{Introduction}

Data duplication is a generic problem in the broad field of data engineering and information systems. There is a lot of affected disciplines that have worked very hard on this topic over the years like, for example, the fields of statistics, databases or artificial intelligence \cite{Elmagarmid2007,Costa2011}. In network traffic monitoring and analysis, data duplication appears as duplicate packets.

Most techniques belonging to this discipline rely on a first step of capturing network packets. There are several ways to capture traffic from Ethernet-based packet-switched networks, but the best suited to meet today's requirements is called port mirroring (\textit{Switched Port Analyzer} or SPAN in Cisco Systems nomenclature). Network hardware, like switches and routers, usually implements this functionality. These devices have the ability to monitor multiple high speed links and send a copy of every packet to a mirror port that can be used to capture traffic.

Manufacturers like Cisco or Hewlett-Packard also allow monitoring Virtual LANs (VLANs) in their equipment. This feature makes monitoring easier and, therefore, it is the most common practice. This mechanism is simply a shortcut to monitor all the ports that belong to a particular VLAN in a single command. Therefore, hereafter we will use \textit{monitored port} to refer indistinctly to both cases, port-based SPAN or VLAN-based SPAN, because in fact they represent the same problem.


\begin{figure}[!t]
	\centering
	\includegraphics[width=.94\linewidth]{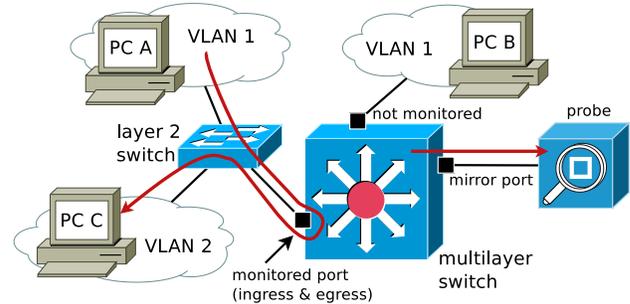}
	\caption{A very simple port mirroring scheme.}
	\label{fig:span}
\end{figure}


Fig.~\ref{fig:span} shows a very simple port mirroring scheme in which a single port is being monitored. PC A in VLAN 1 sends IP packets to PC C in VLAN 2. These packets are routed at the multilayer switch. At the mirror port of the switch, we obtain both the ingress and egress copies from every packet belonging to this stream, which constitute duplicate packets. It is obvious that monitoring only the ingress traffic to this port avoids data duplication, but in such a case, a packet coming from PC B to PC A or C would not be captured.

It's important to note that not all the traffic is duplicated, but only those packets coming to a monitored port and leaving from another monitored port (or the same one). This is the essence of duplicate packets. In Fig.~\ref{fig:span}, the traffic between PCs A-C will be duplicated, but the traffic between PCs B-C and A-B will not.

This simple example shows that, in a practical way, duplicate packets are unavoidable. Every packet traversing two monitored links will appear twice in the network trace. As a result, those duplicates constitute a noise signal that may mislead subsequent analysis in two main ways:

\begin{itemize}
 \item \textbf{Perverting the volume of information}. Packet duplication implies throughput duplication, which impacts on Service Level Agreement (SLA) planning and threshold-based alerting. Moreover, this throughput duplication does not occur in a homogeneous fashion: as has already been mentioned, some traffic may be affected and other may not. This point turns into estimation errors in traffic characterization through the traffic matrix, the identification of heavy hitters or the analysis of packet size distributions, protocol/application mix, etc.
 
 \item \textbf{Complicating the tracking of stateful connections}. For example, duplicate sequence numbers within TCP connections can be mistaken for valid TCP retransmissions.
\end{itemize}


Nevertheless, in spite of its importance, too little attention has been paid to this packet duplication problem. This work is intended to fill this gap. The main contributions of this paper are the analysis of the problem, with a description and classification of the different types of duplicates; the development of a methodology for network packet duplicate detection and removal, and an analytical and experimental dimensioning of this methodology.


The remainder of the paper is organized as follows. Section~\ref{sec:problem} describes the problem of duplicate packets in network traffic monitoring. We discuss several mechanisms that produce distinct types of duplicates. Section~\ref{sec:method} presents a methodology for off-line detection of duplicate packets in order to avoid them in subsequent analysis. Section \ref{sec:results} discusses efficiency aspects. We present some analytical and experimental results and their implications on the fine tuning of our methodology. A dimensioning rule regarding our approach is provided. Finally, Section~\ref{sec:conclusions} summarizes the conclusions of this paper.

\section{Duplicates in network traffic monitoring}\label{sec:problem}

\begin{figure}[!t]
	\centering
	\includegraphics[width=\linewidth]{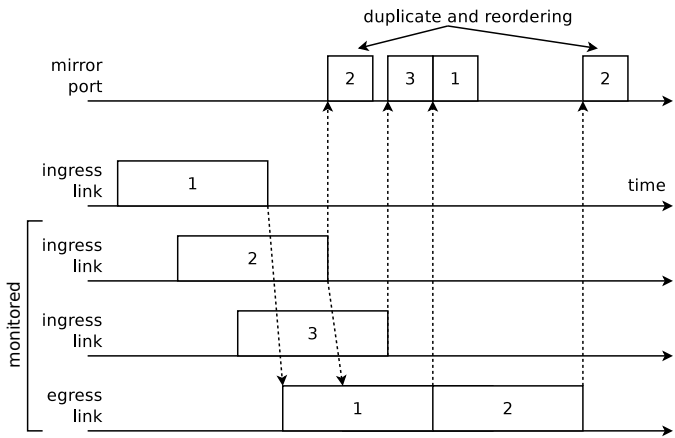}
	\caption{Three packets crossing a device with some monitored ports. The horizontal axis represents time. The mirror port is supposed to be a faster port.}
	\label{fig:duplicates}
\end{figure}

The analysis presented in this section is referred but not limited to a switched Ethernet environment, and the layer 3 is supposed to be IPv4. The IPv6 case is analogous, with the caveat that the IP identification field disappears.

Fig.~\ref{fig:duplicates} shows a schematic of three packets crossing a network device (switch, router...) with some monitored ports:

\begin{itemize}
 \item Packet 1 comes into a not monitored port and goes to a monitored one.
 \item Packet 2 comes into and leaves from a monitored port, therefore it becomes a duplicate at the mirror port.
 \item Packet 3 comes into a monitored port and goes to a not monitored one (not represented).
\end{itemize}

It is assumed that the packets are copied to the mirror after being received or transmitted. Those are the three possible behaviours at any monitored device, and only the second one produces duplicates. Determining which streams will be duplicated becomes challenging in general. Those duplicates consist of an ingress and an egress copy of the same packet. As can be seen in Fig.~\ref{fig:duplicates}, there is a time lag between copies as a consequence of the switching time and the queueing delay, and a variable number of other packets may fall between them.

Moreover, the egress copy can undergo different switching mechanisms that cause distinct types of duplicates. While packet payloads in general remain unchanged, the switching processes (i.e., switching at layer 2, routing, etc.) could lead to several changes in packet headers at various levels of the protocol stack, even when switching at layer 2.

In summary, it is not enough to look for identical packets like some approaches do \cite{editcap}. On the whole, several fields will change, others might or might not change, and others remain the same, like the packet payload. We are going to classify the duplicates depending on the generating mechanism. Table~\ref{tab:types} summarizes the expected changes at different layers.


\begin{table}[!t]
	\renewcommand{\arraystretch}{1.3}
	\caption{Packet modifications for different types of duplicates}
	\label{tab:types}
	\centering
	\begin{tabular}{r|c|l|l}
		\hline
		\bfseries Type & \bfseries Layer & \bfseries Change & \bfseries May change \\
		\hline\hline
		\multirow{3}{*}{Switching} & 2 & & Trunking encap.\\\cline{2-4}
			& \multirow{2}{*}{3} & & DSCP value \\
			& & & Checksum \\
		\hline
		\multirow{4}{*}{Routing} & \multirow{2}{*}{2} & Source address & Trunking encap.\\
			& & Destination address & \\\cline{2-4}
			& \multirow{2}{*}{3} & TTL & DSCP value \\
			& & Checksum & Options \\
		\hline
		\multirow{8}{*}{NAT routing} & \multirow{2}{*}{2} & Source address & Trunking encap.\\
			& & Destination address & \\\cline{2-4}
			& \multirow{4}{*}{3} & TTL & DSCP value \\
			& & Checksum & Options \\
			& & & Source address \\
			& & & Destination address \\\cline{2-4}
			& \multirow{2}{*}{4} & TCP/UDP checksum & Source port \\
			& & & Destination port \\
		\hline
		\multirow{7}{*}{Proxying} & \multirow{2}{*}{2} & Source address & Trunking encap.\\
			& & Destination address & \\\cline{2-4}
			& \multirow{3}{*}{3} & Checksum & DSCP value \\
			& & & Source address \\
			& & & Destination address \\\cline{2-4}
			& \multirow{2}{*}{4} & TCP/UDP checksum & Sequence number \\
			& & & ACK number \\
		\hline
	\end{tabular}
\end{table}

\subsection{Switching duplicates}

They are generated by a switching process at layer 2: the packets enter and leave the device within the same VLAN. Even so, there are some header fields that could change. For example, the egress port could be a trunking port while the ingress not, which implies that egress packets incorporate a 802.1Q VLAN tag at the link layer. In the same manner, the device could be applying QoS policies and thereby the outgoing packets could have different 802.1p priority at Ethernet level or different DSCP value at IP level (wherewith the IP checksum also changes). Note that these changes due to classification and marking could happen for all types of duplicates.

These duplicates can represent up to 50 \% of the traffic captured (even more in particular conditions of packet flooding). They distort MAC-to-MAC stream analysis, hence all the higher layers will be affected. Some equipment, like HP ProCurve switches, are able to automatically suppress these duplicates, but only when the ingress and the egress copies are identical.

\subsection{Routing duplicates}

They are generated by a switching process at layer 3 (or routing). This was the example shown in Fig.~\ref{fig:span}. In this process, some alterations must occur:

\begin{itemize}
 \item The source MAC address is replaced with the device address.
 \item The destination MAC address is replaced with the next hop address.
 \item The Time To Live (TTL) value is decreased by one.
 \item The IP checksum is recalculated.
 \item The IP options may change.
\end{itemize}

\subsection{NAT Routing duplicates}

If the router from the previous case is also working as a Network Address Translation (NAT) device, additional changes are expected to occur \cite{rfcnat}:

\begin{itemize}
 \item The source IP address is replaced with the external address if the packet is leaving the private network. In the opposite direction, the destination IP address is replaced with the internal address.
 \item Both IP addresses are included in the pseudo-header used for IP and TCP/UDP checksum calculation. Therefore, they need to be recalculated.
\end{itemize}

This behaviour constitutes a basic NAT, but it can be extended with port translation:

\begin{itemize}
 \item The source TCP/UDP port is replaced with a mapped port if the packet is leaving the private network. In the opposite direction, the destination TCP/UDP port is replaced with the original port.
\end{itemize}

\subsection{Transparent proxying duplicates}

Transparent proxies and reverse proxies (load balancers) are not uncommon. These devices serve as a basic NAT with transport layer rewriting ability, a technique called \textit{delayed binding} or \textit{TCP splicing} \cite{Syme2003}:

\begin{itemize}
 \item The TCP sequence number is rewritten if the packet is travelling from client to server. In the opposite direction, the TCP ACK number is rewritten.
\end{itemize}

\subsection{Other rare cases}

Today's network infrastructures are dotted with heterogeneous devices that may give rise to unexpected behaviours. Many of these behaviours may respond to, or be consequence of, security reasons. For example, a NAT device effectively isolates a private network from incoming connections, but it also malfunctions with special protocols that carry connection information (IP, port) at application level, such as File Transfer Protocol (FTP) in active mode. Therefore, some of these NATs work as Application Layer Gateways (ALGs) and are able to rewrite the payload \cite{rfcnat2}.

Some of these rare cases are unusual and require special treatment. In the present state, our methodology avoids them.



\section{Duplicate detection methodology}\label{sec:method}

Duplicate packets are not in general identical. Although choosing to compare only payloads might be a good first approach (note, however, that there will be many packets without payload), knowing the type of duplicates is very valuable in order to understand the traffic patterns. This information can only be obtained from packet headers. Indeed, most analysis may require full deduplication, but some of them may not:

\begin{itemize}
 \item If we are calculating the utilization factor per VLAN, switching duplicates must be removed, but routing duplicates must not.
 \item Transport level statistics require removing the routing, routing NAT and proxying duplicates too.
 \item Preserving the routing NAT and proxying duplicates allows us to study both sides of the NAT or proxy.
\end{itemize}

Our methodology works off-line on previously saved captures. It relies on a simple sliding-window model. Every packet is compared against the packets in a windowed buffer using policies derived from the previous section. There is a prototype of this methodology coded in C available at Github \cite{infodups}.

\subsection{The sliding window}

\begin{figure}[!t]
	\centering
	\includegraphics[width=2in]{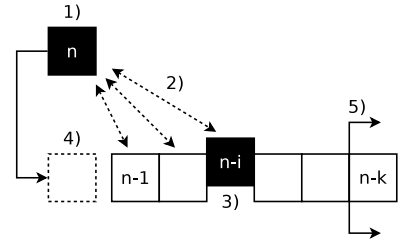}
	\caption{Sliding window for duplicate detection.}\label{fig:window}
\end{figure}

Fig.~\ref{fig:window} shows the sliding window behaviour. The steps are described below:

\begin{enumerate}
 \item With a window of size $k$, the $n$-th packet is read from the trace file.
 \item The $n$-th packet is compared against every packet in the window, from $n-1$ to $n-k$.
 \item If a match is found at the $i$-th comparison, the $n$-th packet is marked as a duplicate from the $n-i$. The search through the window stops.
 \item The $n$-th packet is added to the window.
 \item The window is trimmed to fit a maximum size (older packets are removed).
\end{enumerate}

\subsection{Packet comparison process}

Every packet is dissected in order to determine the highest layer payload that remains unchanged according to our analysis (other protocols might be examined and added):

\begin{itemize}
 \item The TCP/UDP payload.
 \item The IP payload for non-TCP/UDP packets.
 \item The Ethernet payload for non-IP packets.
\end{itemize}

This payload is compared first. If it matches, the header fields are used in order to confirm the identification and decide the type of duplicate. Section~\ref{sec:problem} classifies the distinct types of duplicates and analyzes the expected changes. The rules below must be followed:

\begin{itemize}
 \item \textbf{Fields that do not change}. All of them must be compared.
 \item \textbf{Fields that change}. TTL and checksums are not compared. Source and destination MACs must be compared to ensure that they change.
 \item \textbf{Fields that may change}. Trunking encapsulation, DSCP value and options are not compared. The pairs source/destination IP, source/destination port and ACK/sequence number are compared to ensure that only one changes and the other remains the same.
\end{itemize}

Each type of duplicate has a specific comparator that implements these policies.

\section{Efficiency aspects}\label{sec:results}

\subsection{Efficiency of a single comparison}

According to the previous section, there is a set of fields in a packet that are essential to resolve a duplicate pair. On the other hand, it may be obvious that a pair does not match using a small subset of fields. Therefore, it is important to determine the best order of comparison to avoid wasting valuable time.

The payload constitutes the most significant difference between non-duplicate packets. In order to prove this assumption, we collected a one-day capture at our university's outgoing link. Starting from the deduplicated trace (of around $2 \cdot 10^{9}$ packets), every packet payload was compared against the entire window using the described methodology. For each pair, we collected how many bytes were compared until the mismatch was found.

\begin{figure}[!t]
	\centering
	\includegraphics[width=\linewidth]{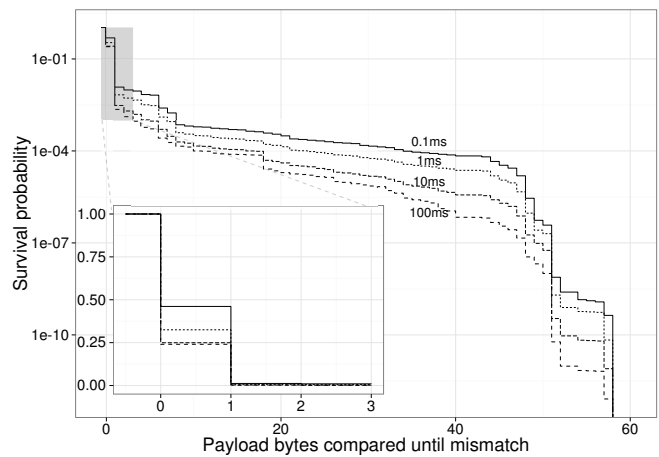}
	\caption{Survival curves for the byte-by-byte payload comparison process with a deduplicated capture. Four window sizes were considered.}\label{fig:survival}
\end{figure}

Four window sizes were tested: 0.1 ms ($\sim 9 \cdot 10^9$ comparisons), 1 ms ($\sim 6 \cdot 10^{10}$ comparisons), 10 ms ($\sim 5 \cdot 10^{11}$ comparisons) and 100 ms ($\sim 5 \cdot 10^{12}$ comparisons). The probability survival curve in Fig.~\ref{fig:survival} shows that, at worst, about 50~\% of non-duplicate pairs were discarded before entering the payload comparison (because protocols or payload sizes didn't match or were equal to zero), and the following 49 \% was discarded within the first two bytes.

This finding states that comparing the payload in the first place, before the header fields, is more efficient in order to discard non-duplicate pairs.

\subsection{Reducing the number of comparisons}

In the worst case, a non-duplicate packet is compared against the entire window. Therefore, using the smallest possible window without compromising the quality of the detection process is desirable because the algorithm complexity scales linearly with this size. This size can be established in terms of time (time-sliding window) or number of packets (element-based sliding window). In order to decide the window size, we are interested in the maximum distance between duplicates.

Previous sections established that a pair of duplicates is formed by an ingress and an egress copy of the same packet. We are going to assume that the switch internally behaves as follows:

\begin{enumerate}
 \item One packet arrives at a monitored port.
 \item The mirror port gets the ingress copy.
 \item The packet is switched to an output queue of another (or the same) monitored port.
 \item After the queue, the mirror port gets the egress copy.
 \item The packet is transmitted.
\end{enumerate}

This behaviour is going to be analytically modeled and experimentally tested.

Between both copies, there is a time spent in the system $s_n$ that can be decomposed into the switching time (or decision time) $x_n$ and the queueing time (or waiting time in the output queue) $w_n$. Applying the expectation operator:

\begin{equation}\label{eq:sn}
\bar{s}=\bar{x}+\bar{w}
\end{equation}

The time spent in the system can be measured as the time difference between the ingress and egress copy of a packet at the mirror port. Within this lag, several copies of other packets might be sent to the mirror port and fall between them, as seen in Fig.~\ref{fig:duplicates}. In our model, the packet difference between duplicates, $\Delta n$, is expected to be proportional to this system time:

\begin{equation}\label{eq:dn}
\bar{\Delta n}=\sum \mu_i \bar{s}
\end{equation}

\noindent where $\mu_i$ is the average service rate at the $i$-th monitored port, what we shall call interfering traffic (i.e., packets 1 and 3 in Fig.~\ref{fig:duplicates}).


This simple model predicts a device-dependent upper bound in terms of time. On the other hand, the number of packets falling between duplicates grows linearly with the device load. In order to test this hypothesis, we conducted an experimental study making use of our implementation \cite{infodups}. Fig.~\ref{fig:scenario} sketches the experimental setup. Three streams are defined:

\begin{figure}[!t]
	\centering
	\includegraphics[width=\linewidth]{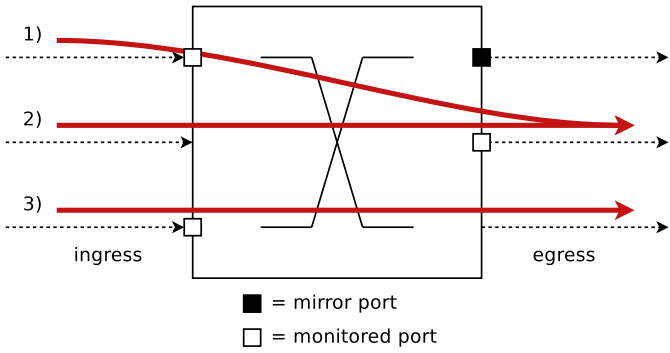}
	\caption{Experimental setup.}\label{fig:scenario}
\end{figure}

\begin{enumerate}
 \item \textit{Main stream}: it is monitored at both the incoming and the outgoing link, so it produces duplicate packets.
 \item \textit{Auxiliary stream}: it is monitored at the outgoing link in the same port that the main stream. It allows us to force queueing at this port.
 \item \textit{Interfering stream}: it is monitored once at the incoming link and lets us insert packets between duplicates at different rates.
\end{enumerate}

Our testbed comprises the following hardware:

\begin{itemize}
 \item One Cisco Catalyst 3560 Switch with 8 FastEthernet ports and 1 Gigabit port (used as mirror port).
 \item Two Dell PowerEdge T110 servers with Intel Xeon X3460 2.8 GHz CPU, 4 GB DDR3 RAM and Intel Gigabit E1G44ET Quad Port PCIe card.
\end{itemize}

The aim of this experiment is to evaluate the impact of the interfering stream into the maximum distance between duplicates. Main and auxiliary streams are formed by maximum size Ethernet packets (PDU $+$ headers $+$ MAC preamble $+$ frame delimiter $+$ CRC $+$ inter-frame gap $=$ 1538 octets) in order to maximize the processing time. The interfering stream is formed by minimum size packets (84 octets), allowing us to sweep a longer range of packets per second. All these streams were generad with exponential interarrival times, making use of the software D-ITG \cite{Botta2012}.

If no losses occur at the output queue, it is equivalent to an infinite queue. Furthermore, if we suppose approximately Poisson arrivals and a deterministic service time, the output queue can be described as a M/D/1. Based on this premise, we can calculate the average queue length \cite{Kleinrock1976}:

\begin{equation}\label{eq:N}
\bar{N}_q = \frac{\rho^2}{2(1-\rho)}
\end{equation}

\noindent and the system time \cite{Kleinrock1976}:

\begin{equation}\label{eq:s}
\bar{s} = \frac{\delta}{2} \left( \frac{2-\rho}{1-\rho} \right)
\end{equation}

\noindent where $\rho$ is the utilization factor and $\delta$ is the service time. The default maximum length of the output queue for our Cisco 3560 switch is 40 packets. In order to avoid losses, we target an average queue length of 5 packets. Thus, Equation (\ref{eq:N}) with $\bar{N}_q = 5$ gives a utilization factor $\rho \approx 0.916$, or 91.6 of 100 Mbps (45.8 Mbps, or 3722 pps, per stream). A greater utilization factor increases the system time. Therefore, we are working with a moderately bad case.

The deterministic service time can be obtained as follows:

\begin{equation}\label{eq:d}
\delta = \frac{L}{C}
\end{equation}

\noindent where $L$ is the length of packets for the main stream ($L=1538\cdot 8$ bits) and $C$ is the link capacity ($C=100$ Mbps). With these values, Equation (\ref{eq:s}) predicts an average system time $\bar{s} \approx 0.79$ ms.

\begin{table}[!t]
	\renewcommand{\arraystretch}{1.3}
	\caption{Interfering stream rates with 84 byte packets.}
	\label{tab:pps}
	\centering
	\begin{tabular}{rrr}
		\hline
		\bfseries Rate (pps) & \bfseries Rate (Mbps) \\
		\hline\hline
		5000 & 3.36 \\
		25000 & 16.80 \\
		45000 & 30.24 \\
		65000 & 43.68 \\
		85000 & 57.12 \\
		105000 & 70.56 \\
		125000 & 84.00 \\
		\hline
	\end{tabular}
\end{table}

\begin{figure}[!t]
	\centering
	\includegraphics[width=\linewidth]{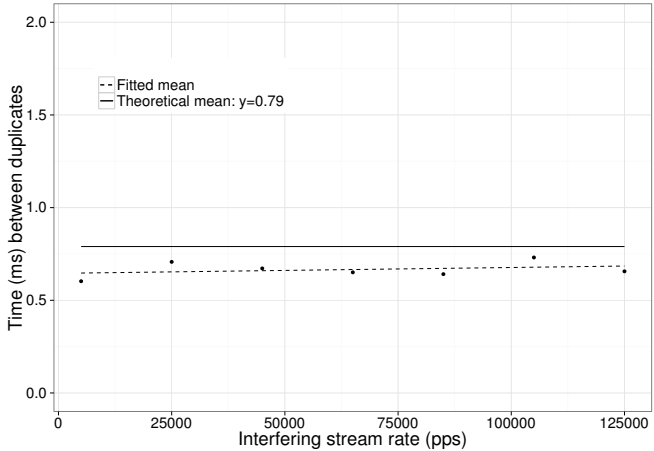}
	\caption{Average time difference between duplicates.}\label{fig:mean-time}
\end{figure}

\begin{figure}[!t]
	\centering
	\includegraphics[width=\linewidth]{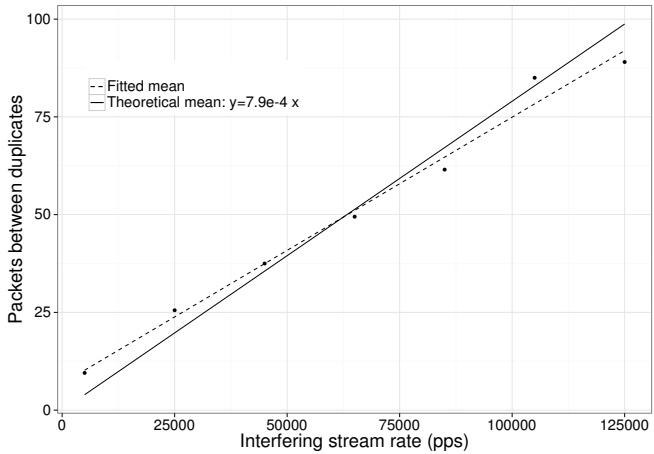}
	\caption{Average number of packets between duplicates.}\label{fig:mean-packets}
\end{figure}

Different interfering stream rates were used, as shown in Table \ref{tab:pps}. For each case, a traffic capture was collected and analyzed so as to extract time differences and number of packets between packet pairs comprising duplicates. Fig.~\ref{fig:mean-time} compares the measured average time difference with the theoretical value. As expected, our model predicts the mean time spent in the system. It is constant and independent from the interfering stream rate. Fig.~\ref{fig:mean-packets} shows the linear growth predicted in the average number of packets falling between duplicates as the interfering stream rate increases.


%
%

The number of packets falling between duplicates depends as much on the system time as on the packet arrival rate of the interfering traffic. Meanwhile, the time difference depends only on the system time. As a result, a time-sliding window becomes the best strategy.

Furthermore, it can be assumed that the switching time is negligible as compared to the queueing time ($x_n<<w_n$). This fact suggests that an upper limit for the system time could be established as follows:

\begin{equation}\label{eq:max}
 max(s_n) \approx max(w_n) = \frac{max(N_q) \cdot max(M)}{C}
\end{equation}

\noindent where $max(N_{q})$ is the maximum length of the output queue, $max(M)$ is the maximum length of a packet in that port and $C$ is the link capacity. With our testbed, $max(N_q)=40$, $max(M)=1538\cdot 8$ bits, $C = 100$ Mbps and $max(s_n) \approx 5$~ms.

On the other hand, this study does not take into account the potential effect of a congested mirror port. The reason is that this scenario would have a harmful effect into a monitoring system because of losses, and therefore it needs to be avoided. In our experiments, the mirror port deals with an utilization factor of about 0.2 (200/1000 Mbps), which means that there can be an additional queueing delay.

In summary, the time difference between duplicates involves three contributions: the switching time $x_n$, the queueing time at output port $w_n$ and the queueing time at mirror port $w'_n$. Being very cautious and assuming that $x_n \leq w_n$ and $w'_n \approx w_n$, Equation (\ref{eq:max}) can be modified to infer a time-based upper limit for the window size ($WS$):

\begin{equation}
 WS = \frac{3 \cdot max(N_{q}) \cdot max(M)}{min(C)}
\end{equation}

\noindent where $max(N_{q})$ is the maximum length of the largest queue, $max(M)$ is the maximum length of a packet and $min(C)$ is the slowest link capacity. In our case, a window size of $WS~=~15$~ms would be advisable. In fact, the maximum separation observed was 8.3~ms.

\subsection{Towards an on-line detection methodology}

This work proposes an off-line methodology. However, many monitoring tools perform on-line traffic analysis without saving network packets to disk. Moreover, one of the most harmful effects of duplicate packets is that they consume a lot of bandwidth at the mirror port. Accordingly, it seems clear that it would be desirable to include an on-line methodology in network devices. Unfortunately, making searches over a sliding window are a very heavy task. Thus, even with parallelization, on-line detection becomes challenging.

Nevertheless, as mentioned in a previous section, duplicates emerge because of a fixed configuration and, therefore, they occur in a deterministic way. It should be possible to find an algorithm that learns how duplicates are generated. After the learning phase, this algorithm could be able to remove duplicates without further analysis. In this way, an on-line detection methodology would be feasible and needs to be further investigated.

\section{Conclusions}\label{sec:conclusions}

This paper addresses an important and unattended problem concerning network traffic monitoring activities. The theoretical background has been exposed: generating mechanisms, types of duplicates and their characteristics are described. Within this context, a detection methodology is proposed, which shall serve as reference for future works.

Moreover, an analytical and experimental study has been conducted regarding the fine tuning of this methodology. As a result, the use of a time-sliding window is recommended. The dimensioning rule provided slightly overestimates the maximum distance between duplicates. Thus, further research with other equipment is needed in order to refine this result.

\bibliographystyle{IEEEtran}
\bibliography{duplicates}
\end{document}